\documentstyle[preprint,aps]{revtex}
\draft
\begin{document}               
\title{
On ``Do the attractive bosons condense?'' by
\\
N. K. Wilkin, J. M. F. Gunn, 
R. A. Smith.
}
\author{M. S. Hussein and  O.K. Vorov
}
\address{
Instituto de Fisica, 
Universidade de Sao Paulo \\
Caixa Postal 66318,  05315-970,  \\
Sao Paulo, SP, Brasil
}
\date{6 August 2001}
\maketitle
\begin{abstract}
Using Perron-Frobenius theorem, we prove that the results
by Wilkin, Gunn and Smith \cite{WGS}
for the ground states of  $N$ Bose atoms rotating at the
angular momentum $L$ in a harmonic
atomic trap with frequency $\omega$ 
interacting via attractive $\delta$$^2$$($$r$$)$ forces,
are valid for a broad class of predominantly attractive
interactions $V(r)$, not necessarily attractive for any $r$.
The sufficient condition for the interaction is that all
the two-body matrix elements 
$\langle$$\bar{z_1}^k$$\bar{z_2}^l$$|$$V$$|$$ z_2^m$$z_1^n\rangle$
allowed by the conservation of angular momentum $k+$$l$$=$$m$$+$$n$,
are negative.
This class includes, in particular, the Gaussian attraction 
of arbitrary radius, 
$\frac{-1}{r}$-Coulomb and $log$$($$r$$)$-Coulomb forces, as
well as 
all the short-range $R$$\ll$$\omega$$^{-1/2}$
interactions satisfying inequality $\int d^2 \vec{r} V(r) < 0$.
There is no condensation at $L\gg 1$, and the angular momentum 
is concentrated in the collective ``center-of-mass'' mode.
\end{abstract}
\pacs{PACS numbers: 03.75.Fi,
32.80.Pj, 67.40.Db, 03.65.Fd
}

Recently,  Wilkin, Gunn and Smith 
have 
shown in Ref.\cite{WGS}  that
the minimum energy states at given angular momentum $L$
of a system of $N$ bosons in a 2d spherically symmetric harmonic
trap with frequency $\omega$$=1$, interacting 
via {\it attractive} 
$\delta$$^2$$($$r$$)$-forces
have,
in the weak coupling limit, 
the form 
\begin{eqnarray}\label{PSI}
\Psi_L= G\left(\sum_{k=1}^{k=N} z_k \right)^L,
G=e^{-\sum\limits_{i=1}^{i=N}\frac{|z_i|^2}{2}}, 
z_k=x_k+iy_k
\end{eqnarray}
These non-degenerate ground states reveal no condensation
at large $L$\cite{WGS}.
Here
we
show that their result is valid for a broad
class of 
forces, predominantly attractive, but not necessarily everywhere
attractive.

We reformulate
the arguments of Ref.\cite{WGS} in the way that they can be viewed as a
particular case of 
Perron-Frobenius theorem 
\cite{BOOK-MATRICES}: 
if a matrix $M$$_{\alpha\beta}$ is 
irreducible {\bf (A)} 
and its 
entries are non-negative {\bf (B)},
then
its positive eigenvector $\sum$$c$$_{\alpha}$$|$$\alpha$$\rangle$ 
(all $c$$_{\alpha}$$>$$0$) 
has {\it maximum eigenvalue}, which is non-degenerate.
Irreducibility means
that there is a chain
$|$$1$$\rangle$$\rightarrow$$|$$2$
$\rangle$$\rightarrow$$.$$.$$.$$|$$p$$\rangle$,
connecting all the basis states, such that 
$M$$_{\alpha,\alpha+1}$$\neq$$0$
for any 
$|$$\alpha$$\rangle$$\rightarrow$$|$$\alpha$$+$$1$$\rangle$.
In this case, the matrix can not be expressed
in block-diagonal form by means of permutations 
of rows and columns.

In the weak coupling limit, 
the Hilbert space ${\cal H}$
of the problem
is spanned by the 
vectors \cite{WGS}
\begin{eqnarray}\label{BASIS}
| \alpha \rangle \equiv [ l _1
, l _2 , . . , l _N ]
\equiv G P _S z_1^{l_1}
z_2^{l_2} . . . z_N^{l_N}, \qquad 
\sum\limits_{i=0}^{N} l _i = L.
\end{eqnarray}
Each vector corresponds to a given partition of integer $L$.
Here,  
$P_S$ denotes symmetrization.
The state 
$\Psi$$_L$$=$$\sum$$c$$_{\alpha}$$|$$\alpha$$\rangle$(\ref{PSI})
has all $c$$_{\alpha}$$>$$0$ in this basis\cite{WGS},
and 
$\Psi$$_L$
is eigenstate of any interaction
$\sum_{i>j} V(|\vec{r}_i-\vec{r}_j|)$
in ${\cal H}$\cite{HV3}
with 
eigenvalue
\begin{displaymath}
\frac{1}{2}(N^2-N)\int_0^{\infty}
rdre^{-r^2/2}V(r).
\end{displaymath}
Therefore, the state 
(\ref{PSI})
must be 
non-degenerate ground state
of
any interaction $V$, whose matrix 
$M$$_{\alpha,\beta}$
$=$$-$$V$$_{\alpha,\beta}$ 
obeys the conditions 
of the Perron-Frobenius theorem, 
{\bf (A)} and {\bf (B)}. 
Since operation 
$M$$_{\alpha\alpha}$$\rightarrow$$M$$_{\alpha\alpha}$$+$$const$
does not affect eigenvectors, {\bf (B)} 
reads $M$$_{\alpha\neq\beta}$$\geq$$0$.
The
off-diagonal
matrix elements 
are given by
\begin{eqnarray}\label{matrix}
-M_{\alpha\beta}=V_{\alpha\beta}=\sum_{klmn(m>l)}
S_{kl,mn}^{\alpha,\beta}
V_{kl,mn}, \quad
S_{kl,mn}^{\alpha,\beta} \geq 0,
\end{eqnarray}
with 
\begin{displaymath}
 V_{kl,mn}  \equiv 
 \langle  \bar{z_1}^k  \bar{z_2}^l  V  z_2^m  z_1^n  \rangle  
 =  
\int d^2 z_1 \int d^2 z_2  \bar{z_1}^k  \bar{z_2}^l
V\left(\sqrt{|z_1-z_2|^2}\right)
 z_2^m  z_1^n Q
=\delta_{k+l,m+n}  {\cal V}  _{klm}  [  V  ] 
\end{displaymath}
the two-body matrix element
obeying
conservation of the angular momentum. Here, 
$Q=\frac{e^{-|z_1|^2-|z_2|^2}}{\pi^2\sqrt{k!l!m!n!}}$,
bar denotes 
complex conjugation and 
$S$ are some non-negative
quantities [Cf.(\ref{BASIS})].
Now,
we show that the conditions 
\begin{equation}\label{CONDITION2body}
{\cal V}_{klm}[V] < 0
\end{equation}
$($$l$$>$$m$)
are sufficient for both {\bf(A)} and {\bf(B)}. 
Indeed, the connecting chain 
for the basis states (\ref{BASIS})
\begin{displaymath} 
[  L  ,  0  ,  0,  .  .  ]  \rightarrow 
 [  L  -  1,  1  ,  0  ,  .  .  ] 
 \rightarrow  [  L  -  2  ,2  ,  0  ,  .  .  ] 
 \rightarrow  [  L-  2  , 
 1  ,  1,  .  .  ]\rightarrow  .  .  \rightarrow 
 [  1  ,  1  ,  .  .  1  .  .  .  ]
\end{displaymath}
is found keeping the only terms
$V$$_{kl,l-1 k+1}$ in 
(\ref{matrix}).
Adding other
$V$$_{kl,mn}$ can produce
no cancellations 
by virtue of (\ref{CONDITION2body}).
For the same reason, all 
$M$$_{\alpha\beta}$$\geq$$0$.

In particular, the conditions (\ref{CONDITION2body}) hold
for
the attractive $\delta$-function, as we have 
\begin{displaymath}
 {\cal V}  _{klm}  (  -  \delta  )  = 
 \frac{-(k+l)!2^{-(k+l+1)}}{\pi\sqrt{k!l!m!(k+l-m)!}} .
\end{displaymath}

In general case, (\ref{CONDITION2body}) reads
\begin{eqnarray}\label{CONDITION-ME}
{\cal V}_{klm}  = 
\sum\limits_{i,j=0}^{i=m,j=k}
\frac{ a (-2)^{-(i+j)}(\Delta+i+j)!f}
{(m-i)!(\Delta+i)!(k-j)!(\Delta+j)!i!j!} < 0,
\end{eqnarray}
where 
$a$$=$$\frac{\sqrt{k!l!m!(k+\Delta)!}}{2^{l-m}}$ and 
$\Delta$$=$$l$$-$$m$$>$$0$, the function $f$ is defined by 
\begin{displaymath}
 f  =  \int_0^{\infty}  r  
d  r  V  (  r  )  L_{\Delta+i+j}  (  \frac{r^2}{2}  )  e^{\frac{-r^2}{2}}
\end{displaymath}
with $L_K(x)$ the Laguerre polynomial\cite{ABR}.

For short-range interactions $V$$_R$$($$r$$)$ with effective radius
$R$  
much smaller than the oscillator
length, $R$$\ll$$1$,
(\ref{CONDITION-ME})
is reduced to
\begin{displaymath}
{\cal V}_{klm}(V_{R\ll1}) = {\cal V}_{klm}(\delta)
\int d^2\vec{r} V_{R\ll1}(r)
\end{displaymath}and 
the sufficiency conditions  
(\ref{CONDITION2body},\ref{CONDITION-ME})
are therefore replaced by the single inequality
\begin{displaymath}
\int d^2 \vec{r} V_{R\ll1}(r) < 0.  
\end{displaymath}
Thus, the results (\ref{PSI})
hold for
short-range 
($R$$\ll$$1$$\equiv$$\sqrt{\frac{\hbar}{m\omega}}$), 
interactions
which 
are attractive on average.

The conditions  (\ref{CONDITION2body}) can be seen valid
for
a wide class of long-range interactions.
The Gaussian attractive interaction, 
$V=\frac{ -e^{\frac{-r^2}{R^2}} }{\pi R^2}$,
gives
\begin{eqnarray}\label{GAUSSIAN}
{\cal V}_{klm}\left[\frac{ -e^{\frac{-r^2}{R^2}} }{\pi R^2}\right]=
\frac{-1}{{\pi}\Delta!}\sqrt{\frac{l!(k+\Delta)!}{k!m!}}
\frac{ (1+R^2)^{k+m}}{ (2+R^2)^{k+l+1}}
\quad  {_{2}}  F  _{1}  [  -  k  , 
 -  m  ;  l  -  m  +  1  ; 
 \frac{1}{(1+R^2)^2}  ] 
< 0 ,
\end{eqnarray}
where ${_{2}}  F  _{1}  [ a  , b  ;  c  ; x  ]$ 
is the hypergeometric function\cite{ABR}
which
is seen to be
positive 
for any $R$\cite{ABR}.
The results for log-Coulomb forces, $V=log(r)$, and Coulomb forces,
$V=-1/r$,
can be obtained from (\ref{GAUSSIAN}). We have
\begin{displaymath}
 {\cal V}  _{klm} 
 [  l  o  g  (  r  )  ]  = 
 \frac{\pi}{2}  \int_0^{\infty} 
 d  (R  ^2)  {\cal V}  _{klm}  \left[  \frac{-e^{-r^2/R^2}}{\pi R^2}  \right]  
\end{displaymath}
and 
\begin{displaymath}
 {\cal V}  _{klm}  \left[  \frac{-1}{r}  \right]  = 
 2  \sqrt{\pi} 
 \int_0^{\infty}  d  R  {\cal V}  _{klm}  [  \frac{-e^{-r^2/R^2}}{\pi
R^2}  ] ,
\end{displaymath}
respectively. It is seen that ${\cal V}$$_{klm}$$\leq$$0$ in both
cases.

We conclude therefore 
that the results (\ref{PSI}) for the 
``yrast states'' of weakly attractive Bose atoms
in harmonic trap, derived by Wilkin, Gunn and Smith in Ref.\cite{WGS} 
for the case of $\delta$-forces, are valid for a wide class 
of interactions which are predominantly attractive 
(\ref{CONDITION2body},\ref{CONDITION-ME}):
There is no condensation at $L\gg 1$(\ref{PSI}), and the angular momentum 
is concentrated in the collective ``center-of-mass'' mode.

This work was supported by FAPESP.

\end{document}